\documentclass[letterpaper,twocolumn,10pt]{article}
\usepackage{usenix2019_v3}

\usepackage{graphicx}
\usepackage{subcaption}
\usepackage{booktabs}
\usepackage{multirow}
\usepackage{array}
\usepackage{placeins}
\usepackage{siunitx}
\usepackage{tikz}
\usetikzlibrary{positioning,arrows.meta,fit,calc,shapes.geometric,shapes.misc,patterns,decorations.pathreplacing,backgrounds}
\usepackage[most]{tcolorbox}

\usepackage{amsmath,amssymb}
\usepackage{etoolbox}
\robustify{\nobreakdash}
\usepackage{amsthm}

\usepackage{listings}
\usepackage{enumitem}
\setlist{nosep,leftmargin=*}
\usepackage{xspace}
\usepackage[capitalise,noabbrev]{cleveref}

\hypersetup{pdftitle={Pin and Replay, or Refuse: Conditional Exactness and Probabilistic Validation},colorlinks=true,linkcolor=blue!50!black,citecolor=blue!50!black,urlcolor=blue!50!black}

\theoremstyle{definition}

\newtheorem{proposition}{Proposition}

\definecolor{cStable}{HTML}{9ECAE1}
\definecolor{cEdited}{HTML}{FD8D3C}
\definecolor{cNew}{HTML}{E6550D}
\definecolor{cTare}{HTML}{2171B5}
\definecolor{cReplay}{HTML}{6BAED6}
\definecolor{cOk}{HTML}{31A354}
\definecolor{cBad}{HTML}{CB181D}
\definecolor{cGrey}{HTML}{636363}

\newtcolorbox{plainbox}[1]{enhanced,breakable,colback=cTare!4,colframe=cTare!70!black,boxrule=0.5pt,arc=1.5pt,
  left=4pt,right=4pt,top=3pt,bottom=3pt,fonttitle=\bfseries\small,title={#1},fontupper=\small}

\newcommand{\sys}{TARE\xspace}
\newcommand{\sysexpansion}{Tool Annotation Replay with Evidence}
\author{%
{\rm Joshua Terranova}\\
Thomas Lord Department of Computer Science\\
University of Southern California
\and
{\rm Jiapeng Zhang}\\
Thomas Lord Department of Computer Science\\
University of Southern California%
}
\hypersetup{pdfauthor={Joshua Terranova, Jiapeng Zhang}}

\newcommand{\dec}[1]{\textsc{#1}}
\newcommand{\cutga}{\texttt{-{}-cut\_ga}\xspace}

\newcommand{\HeadlineBoundedMedian}{3.49}
\newcommand{\HeadlineBoundedCIlow}{3.32}
\newcommand{\HeadlineBoundedCIhigh}{4.10}

\newcommand{\HeadlineMedian}{3.47}
\newcommand{\HeadlineCIlow}{3.30}
\newcommand{\HeadlineCIhigh}{3.81}
\newcommand{\HeadlineN}{8}
\newcommand{\HeadlineGeoMean}{3.51}
\newcommand{\HeadlineRangeLow}{3.28}
\newcommand{\HeadlineRangeHigh}{3.96}
\newcommand{\FlaggedCount}{1}
\newcommand{\FlaggedList}{GCF\_049949335.1}
\newcommand{\SensitivityAllGenomes}{3.44}

\newcommand{\CostOfCheckingMedian}{1.58}

\newcommand{\CostOfCheckingRangeLow}{1.53}
\newcommand{\CostOfCheckingRangeHigh}{1.71}

\newcommand{\CampaignBodyLines}{466{,}725}

\newcommand{\SgA}{3.44}
\newcommand{\CBA}{1.54}
\newcommand{\SpreadA}{4.6}

\newcommand{\SgB}{3.28}
\newcommand{\CBB}{1.71}
\newcommand{\SpreadB}{2.9}

\newcommand{\SgC}{3.81}
\newcommand{\CBC}{1.66}
\newcommand{\SpreadC}{40.3}

\newcommand{\SgD}{3.28}
\newcommand{\CBD}{1.57}
\newcommand{\SpreadD}{48.0}

\newcommand{\SgE}{3.50}
\newcommand{\CBE}{1.53}
\newcommand{\SpreadE}{10.1}

\newcommand{\SgF}{3.22}
\newcommand{\CBF}{1.61}
\newcommand{\SpreadF}{55.5}

\newcommand{\SgG}{3.53}
\newcommand{\CBG}{1.57}
\newcommand{\SpreadG}{44.8}

\newcommand{\SgH}{3.31}
\newcommand{\CBH}{1.58}
\newcommand{\SpreadH}{3.3}

\newcommand{\SgI}{3.96}
\newcommand{\CBI}{1.64}
\newcommand{\SpreadI}{13.1}

\title{Pin and Replay, or Refuse: Conditional Exactness and\\
Probabilistic Validation Across Reference Releases}
\date{}

\begin{document}
\maketitle

\begin{abstract}
Biologists annotate genomes by searching their proteins against reference databases such as Pfam, which publish a
new release every few weeks to months. Each release forces a full re\nobreakdash-search, although most of the database has not
changed: 86.6\% of Pfam's entries matched between releases 38.1 and 38.2 after excluding build metadata and comments.
Reusing the old results is hard, because the search programs are fixed binaries whose printed statistics depend on
the size of the whole database. We present \sys (\sysexpansion), a wrapper that reruns an unmodified
program only on the entries that can matter, with the program's own database\nobreakdash-size setting pinned to the new
release, so that the program itself prints the numbers a full run would print. \sys then tests the replay by adding
back entries it left out, and refuses with a failure status if any line changes, dumping a full stock body
when the tool still runs.
Two claims stay separate: \textit{conditional exactness} under a stated contract (tested each update, not proved), and
\textit{runtime assurance} as probabilistic validation at a stated prevalence---not a guarantee against shipping any
incorrect output.
In the all\nobreakdash-record timing, \sys also audits 299 records drawn from all queries, so a hidden dependency affecting at
least 1\% of that population ships with probability at most 5\% (valid oracle); rarer faults are outside that bound.

On Pfam 38.1$\rightarrow$38.2, \sys is $\HeadlineMedian\times$ faster than a full rerun under that all\nobreakdash-record bounded audit
(median over \HeadlineN\ genomes after excluding \FlaggedCount\ wall\nobreakdash-quality flagged genome; 95\% CI
\HeadlineCIlow--\HeadlineCIhigh). Historical candidate\nobreakdash-population timings on the same nine genomes were $4.24\times$
(Bernoulli audit; 95\% CI 4.17--4.68) and $\HeadlineBoundedMedian\times$ (bounded $s{=}299$; 95\% CI
\HeadlineBoundedCIlow--\HeadlineBoundedCIhigh). Cost of checking versus unchecked replay is $\CostOfCheckingMedian\times$
(C/B over the same unflagged set).
Across 715{,}704 output lines from HMMER and Infernal, not one differed from a full run. \sys refused DIAMOND,
whose first filter reads bytes of the neighboring database entry, and BLAST+, whose unpinned probe rows fit none of
\sys's column rules.
\end{abstract}

\section{Introduction}
\label{sec:intro}

When a laboratory sequences a genome, it obtains thousands of protein sequences with no labels. To learn what
each protein does, researchers compare it against \textit{reference databases}: curated collections of known
protein families, such as Pfam~\cite{pfam2025}, UniProt~\cite{uniprot2025}, and InterPro~\cite{paysanlafosse2023interpro}.
A program such as HMMER~\cite{eddy2011hmmer} scores every protein against every family and reports the good
matches, called \textit{hits}. This step, called \textit{annotation}, underlies much of modern biology, from
comparative genomics to the sequence alignments used by structure predictors such as AlphaFold~\cite{jumper2021alphafold}.

\begin{figure*}[t]
\centering
\includegraphics[width=\textwidth]{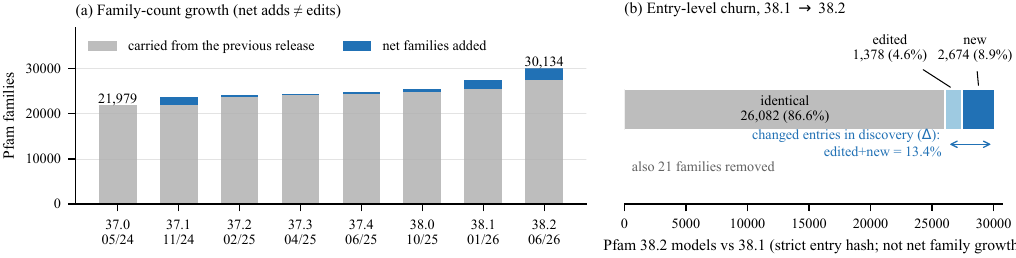}
\caption{\textbf{Reference databases change often, but usually only in part.} (a)~Family\nobreakdash-count growth:
Pfam from 21{,}979 to 30{,}134 over eight releases; blue is \emph{net} families added (not edits).
(b)~Entry\nobreakdash-level strict\nobreakdash-hash comparison of 38.1$\rightarrow$38.2: 86.6\% identical; edited$+$new $=$ 13.4\%,
the changed entries searched during discovery ($\Delta$). Per\nobreakdash-record replay candidates $C_r$ also include
cached hits on stable entries; pads and the audit add further work. Panel~(b) is the timed pair;
\cref{fig:pairs} shows neighboring releases the same way (median $f_\Delta$ 12.2\%).}
\label{fig:churn}
\end{figure*}

Reference databases keep improving. UniProt and InterPro publish a new version, called a \textit{release},
roughly every eight weeks~\cite{uniprot2025,paysanlafosse2023interpro}, and Pfam published eight releases
between May 2024 and June 2026~\cite{pfam382relnotes}. Each release adds and revises families, so annotations
computed against an old release go out of date. The annotation programs keep no memory of earlier runs,
so the only way to update is to compare every protein against the entire new database again. Most of that
work recomputes answers that did not change: between Pfam 38.1 and 38.2, 86.6\% of the 30{,}134 families were
identical apart from build\nobreakdash-date, build\nobreakdash-command, and comment lines (\cref{fig:churn}). At scale this repeated cost shapes practice. NCBI re\nobreakdash-annotates its
RefSeq collection on a rolling schedule because redoing all of it at once ``requires securing at once a
significant amount of computing resources''~\cite{li2021refseq}, and JGI states that at over 65 billion genes
``it is impossible to upgrade all genome and metagenome annotations to the latest version''~\cite{chen2021img}.

\textbf{Why reuse is hard.} It might seem easy to search only the new and edited families and keep the old results
for the rest. It is not, for three reasons. First, the results are \textit{coupled}: the statistic HMMER prints for
each hit, its E-value, is scaled by the total number of families in the database, much as a grade curved to class
size changes for every student when new students enroll. Adding families therefore changes the printed numbers of
hits on families that did not change. Rerunning HMMER on exactly the right families, without accounting for this,
changes 9{,}083 of 9{,}094 output lines for one bacterial genome.
Second, the programs are unmodified binaries
maintained by others; we cannot change their code. Third, an approximately correct answer is not good enough:
pipelines that track provenance must reproduce a full run exactly, because ``pipeline differences can lead to
annotation discrepancies''~\cite{chen2021img}.

Existing tools fall short in two ways. General\nobreakdash-purpose caches, such as build systems and workflow
managers~\cite{curtsinger2022riker,xie2026incr,shiptoski2023processcache,ditommaso2017nextflow,molder2021snakemake},
remember whole commands keyed by the bytes they read. A new release changes the database file, so they rerun the
search in full. Specialized incremental search tools such as iBLAST and iSeqSearch~\cite{dash2021iblast,yoo2025iseqsearch}
are written by hand for one program and adjust the printed numbers after the fact. In the one configuration we ran, only 1.3\% of the rows that iSeqSearch shared
with a full run were byte\nobreakdash-identical, and it reported 133 hits to sequences that the new release had deleted.
\textit{How can we reuse earlier annotations across a database release, without modifying the programs, reprint under a
stated contract when coupling is exposed, refuse when it is not, and state a probabilistic bound on undetected faults
above a prevalence threshold?}

\begin{plainbox}{The key idea in plain terms}
In the programs we studied, entries in the database influence each other's printed results through only a few
global numbers, chiefly the database size, and the programs already accept those numbers as settings. If we tell
the program the new release's size up front (\textit{pin} it), it scores each family on its own merits. Then we can
rerun it on a small subset of the database and it prints exactly what a full run would print, digit for digit,
because the program, not our wrapper, computes every number. This is what \textit{taring} a scale does: fix the
baseline first, then weigh only what is new. When a program breaks this pattern, we try to detect it and fall back to a
full run.
\end{plainbox}

Pinning a global size is not new in itself. Parallel search tools split a database across machines and fix the
database size so that every piece prints consistent E-values~\cite{darling2003mpiblast}, and HMMER's \texttt{-Z} option
exists partly for that purpose~\cite{eddy2011hmmer}. What is missing is a way to use it across releases, for a binary we
do not control, without trusting that the binary has no other coupling. \sys supplies that: it finds which entries
changed (\cref{sec:model}), reruns each protein only against its few candidate families with the size pinned
(\cref{sec:replay}), and before returning anything adds back entries it left out and refuses if any line changes
(\cref{sec:check}). Refusal exits with failure status and writes a full stock body when the tool still runs.
Worked numbers for Pfam 38.1$\rightarrow$38.2: pin \texttt{-Z 30134} and, on one genome, replay 3{,}719 candidate families rather than all 30{,}134 (\cref{fig:pin}).

The primary successful application is HMMER on Pfam: shipped runs under the supported configuration matched full runs,
and the all\nobreakdash-record audit is timed there.
Infernal is supporting evidence from the same tool lineage and obtains its measured speedup under a reduced check that
skips the sampled audit on long\nobreakdash-DNA inputs (\cref{sec:eval-time}).
DIAMOND and BLAST+ demonstrate refusal boundaries; Semgrep ships under full verification but is slower than a stock
rerun.
Our contributions are:

\begin{enumerate}
  \item \textbf{Conditional exactness, separated from probabilistic validation.} Release\nobreakdash-diff selection and
  candidate\nobreakdash-only, batched, pinned replay under a stated contract (\cref{prop:exact}), guarded by a probe gate and
  check\nobreakdash-or\nobreakdash-refuse.
  The audit bounds missing faults that affect at least a stated fraction of the audited population; it does not bound
  shipping any incorrect output (\cref{sec:design}).
  \item \textbf{A root cause for a refusal.} Investigating a failure of our first check revealed why DIAMOND cannot be
  replayed under the contract: its first filter reads bytes belonging to the neighboring database entry, so its results
  depend on database order (\cref{sec:eval-safety}).
  \item \textbf{A pre\nobreakdash-registered HMMER evaluation, with stated boundaries.} Byte agreement over 715{,}704 HMMER and
  Infernal lines; Pfam 38.1$\rightarrow$38.2 all\nobreakdash-record bounded\nobreakdash-audit median $\HeadlineMedian\times$
  (95\% CI \HeadlineCIlow--\HeadlineCIhigh; $\HeadlineN$ unflagged genomes); historical candidate\nobreakdash-population medians
  $4.24\times$ (Bernoulli) and $\HeadlineBoundedMedian\times$ (bounded $s{=}299$); refuse controls; and comparisons with
  caches, approximate incremental search, and GeStore, with designs locked before the measured runs (\cref{sec:eval}).
  Isolation from a competent expert pin script remains unresolved on the locked five\nobreakdash-genome plan
  (\cref{sec:discussion}).
\end{enumerate}

\section{Background}
\label{sec:background}

\begin{table}[t]
\caption{Terms used in this paper.}
\label{tab:glossary}
\centering\footnotesize
\setlength{\tabcolsep}{3pt}
\begin{tabular}{@{}>{\raggedright\arraybackslash}p{1.6cm}>{\raggedright\arraybackslash}p{6.2cm}@{}}
\toprule
Term & Meaning \\
\midrule
Record & One query item, here one protein (or one DNA sequence for Infernal). \\
Entry & One family model in the reference (a profile HMM in Pfam). \\
Release & A new version of the reference. \\
Pin & Fixing a global setting (e.g., $Z$) to the new release's value. \\
Replay & Rerunning the unmodified program on a subset of the reference. \\
\bottomrule
\end{tabular}
\end{table}

\subsection{How Annotation Works}

An annotation program takes two inputs: a set of \textit{records}, such as the thousands of proteins of a bacterial
genome, and a \textit{reference} of \textit{entries}, such as Pfam's 30{,}134 protein families (\cref{tab:glossary}).
Each Pfam family is stored as a statistical profile, a profile hidden Markov model, that summarizes what members
of that family look like. HMMER's \texttt{hmmscan} compares every protein with every family, computes a score, and
writes one line per reported hit, containing the names of the protein and family, the score, and an E-value. The
E-value estimates how many equally good matches would occur by chance in a database of this size, and HMMER computes
it as a probability multiplied by $Z$, the number of families in the database~\cite{eddy2011hmmer}. With Pfam's curated
per\nobreakdash-family score thresholds (the \cutga option), whether a hit is reported depends only on its score against that
family's threshold~\cite{pfam2025}.

\subsection{Why Naive Reuse Fails}

\begin{figure}[t]
\centering
\resizebox{\linewidth}{!}{%
\begin{tikzpicture}[font=\scriptsize, x=1cm, y=1cm]
  \node[anchor=west, font=\scriptsize\bfseries] at (0,2.85)
    {One hit: $E{=}P{\times}Z$ ($P$ fixed by the score)};
  \node[draw, rounded corners=2pt, fill=gray!10, align=center, minimum width=2.1cm, minimum height=0.68cm]
    (P) at (1.15,1.75) {$P{=}3.1{\times}10^{-9}$};
  \node[font=\tiny\color{cGrey}] at (1.15,1.22) {shared};
  \node[font=\Large] at (2.55,1.75) {$\times$};
  \node[draw=cGrey, line width=1.0pt, rounded corners=2pt, align=center,
        minimum width=1.7cm, minimum height=1.45cm, fill=cGrey!7] (Za) at (3.75,1.45) {%
    {\tiny\color{cGrey}cache}\\[1pt]
    {\footnotesize\bfseries $Z{=}27{,}481$}\\[2pt]
    {\scriptsize $E{=}8.5{\times}10^{-5}$}\\[1pt]
    {\scriptsize\bfseries\textcolor{cGrey}{stale}}};
  \node[draw=cBad, line width=1.0pt, rounded corners=2pt, align=center,
        minimum width=1.7cm, minimum height=1.45cm, fill=cBad!7] (Zb) at (5.65,1.45) {%
    {\tiny\color{cGrey}subset}\\[1pt]
    {\footnotesize\bfseries $Z{=}3{,}719$}\\[2pt]
    {\scriptsize $E{=}1.2{\times}10^{-5}$}\\[1pt]
    {\scriptsize\bfseries\textcolor{cBad}{wrong}}};
  \node[draw=cOk, line width=1.1pt, rounded corners=2pt, align=center,
        minimum width=1.7cm, minimum height=1.45cm, fill=cOk!7] (Zc) at (7.55,1.45) {%
    {\tiny\color{cGrey}full/pin}\\[1pt]
    {\footnotesize\bfseries $Z{=}30{,}134$}\\[2pt]
    {\scriptsize $E{=}9.3{\times}10^{-5}$}\\[1pt]
    {\scriptsize\bfseries\textcolor{cOk}{match}}};
  \draw[-{Stealth[length=1.6mm]}, gray!60] (2.9,1.75) -- (Za.west);
  \draw[-{Stealth[length=1.6mm]}, gray!60] (2.9,1.75) -- (Zb.west);
  \draw[-{Stealth[length=1.6mm]}, cOk!80!black] (2.9,1.75) -- (Zc.west);
  \node[draw=cOk!70!black, fill=cOk!6, rounded corners=2pt, anchor=north west,
        text width=8.3cm, align=left, inner sep=3pt] at (0,-0.1)
    {\textbf{Pin:} subset of 3{,}719 with \texttt{-Z 30134} prints the full-rerun $E$ (program rounding).};
\end{tikzpicture}%
}
\caption{\textbf{Why old results cannot simply be reused.} $E{=}PZ$. Keeping the 38.1 cache or searching a
subset without pinning $Z$ changes the printed E-value. Pinning $Z$ to the new release matches the full rerun.
Illustrative $P$; $Z$ from Pfam 38.1/38.2 and one proteome's candidate subset.}
\label{fig:pin}
\end{figure}
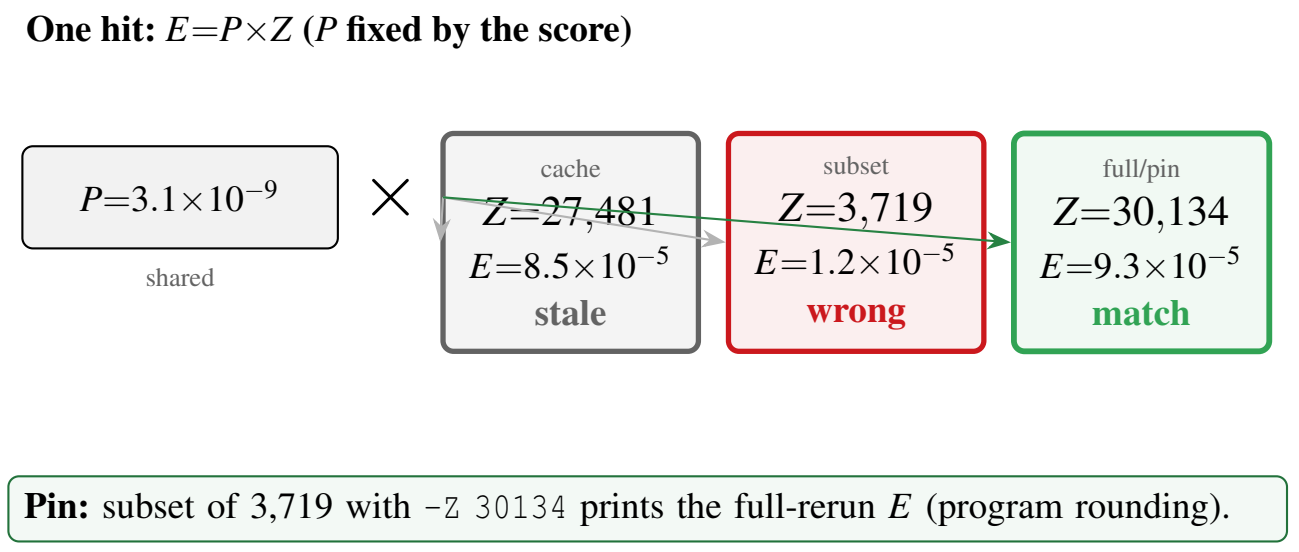

Suppose a laboratory annotated a genome with Pfam 38.1 and wants Pfam 38.2 results. The obvious shortcut is to
search only the 4{,}052 families that are new or edited and keep the cached lines for the other 26{,}082. Nearly
every kept line is then wrong, because its E-value was multiplied by 38.1's $Z$ (27{,}481) instead of 38.2's
(30{,}134) (\cref{fig:pin}). Rerunning HMMER on exactly the right families does not help either, because the program
then uses the subset's size as $Z$: on one genome, such a rerun differed from a full run on 9{,}083 of 9{,}094 lines,
and the other 11 lines matched. Multiplying the cached E-values by
$Z_{\text{new}}/Z_{\text{old}}$ gets closer but still fails, because printed numbers are rounded, and the rounding of
a rescaled number differs from the program's rounding of its own internal value.

\section{Problem Model}
\label{sec:model}

\paragraph{Program and release.}
A black-box program $P$ maps records $R$, reference $E$, and options $o$ to a
tabular body (headers ignored). Each entry of the ordered list $E$ has a strict
hash $h(e)$ (SHA-256 after dropping caller-declared volatile tags). A release
$E\!\to\!E'$ yields stable $S=\{e\in E':h(e)\in h(E)\}$, changed
$\Delta=E'\setminus S$, and $f_\Delta=|\Delta|/|E'|$.
For each pair $(r,e)$, let $f(r,e,o,g(E'))$ be its possibly empty sequence
of emitted body rows, including bytes and multiplicities, where $g(E')$
collects reference-wide quantities (entry count, length, \ldots).
Exact re-annotation reproduces $\mathrm{body}(P(R,E',o))$ under the
conditions below; an observed check failure refuses with failure status
and dumps a full-run body when the tool still runs.

\paragraph{Coupling taxonomy.}
\textbf{C0} (entry-local): $f$ independent of $g$ and of every $e'\neq e$
(Infernal's no-clan variant) $\rightarrow$ replay.
\textbf{C1} (reference-size): $f$ depends on $E'$ only through a normalizer in
$g$ that $P$ exposes or \sys rescales (HMMER \texttt{-Z} for per-sequence
E-values; conditional per-domain E-values use \texttt{-{}-domZ}, recovered under
\texttt{-{}-cut\_ga} by reconstructing the reported hit set) $\rightarrow$ pin, or
refuse if uncovered.
\textbf{C2} (neighbor/order): the line for $(r,e)$ depends on neighbors of $e$
or on input order (DIAMOND) $\rightarrow$ refuse.
\textbf{C3} (cross-record global, e.g.\ FDR) and \textbf{C4} (nondeterministic)
$\rightarrow$ refuse when observed.
Refusal rules R1--R6 form one surface: R1 no contract; R2 size-dependent, no
pin; R3 in-situ pin check fails; R4 non-candidate pair in replay; R5 pad changes
a line; R6 audit differs. C0/C1 replay; C2--C4 refuse (R1--R3 on the probe,
R4--R6 on the update). Side-file rewrites and invalidate-groups are
\textit{declarations}; they do not establish entry locality for competing groups.

\paragraph{Exactness (C0/C1).}
\label{asm:det}\label{asm:pin}\label{asm:key}\label{asm:order}%
Assumptions: (D)~\textbf{determinism}---fixed $(r,e,o,g)$ emits the same rows;
(P)~\textbf{complete pin}---pinned options, together with quantities exactly
reconstructed from the candidate set (e.g.\ HMMER's conditional domain
normalizer under \texttt{-{}-cut\_ga}), fix every reference-wide quantity
affecting row existence, multiplicity, or bytes to its full-release value,
including quantities not observed by the fitter, without changing other
output-relevant behavior;
(K)~\textbf{attribution}---each row occurrence is attributable to its record
and entry, preserving repeated occurrences;
(O)~\textbf{exact assembly}---the merge reconstructs stock ordering and
formatting from these rows and the original record/reference order;
(C)~\textbf{candidate completeness}---every entry producing a full-release
row for $r$ belongs to its candidate set $C_r$;
(I)~\textbf{record-subset invariance}---with reference and options fixed,
running on a subset of records produces exactly the projection of the
full-record body onto that subset.
\begin{proposition}[Exact merge]\label{prop:exact}
Assume C0 or C1 and (D), (P), (K), (O), (C), and (I).
Running $P$ on each $r$ against $C_r$ with $g$ pinned, then applying
the merge in (O), yields $\mathrm{body}(P(R,E',o))$.
\end{proposition}
\begin{proof}
By (I), isolating $r$ preserves its full-reference rows.
Under C0/C1 and (P), restricting the reference to $C_r$ preserves the
emitted rows for each retained pair, including empty sequences and
multiplicities; (D) excludes execution drift. By (C), every omitted pair
emits no full-release row. Replay therefore produces exactly the
full-release row occurrences; (K) preserves attribution and (O)
reconstructs the full body.
\end{proof}
The probe (R1), pin check (R3), pads (R5), and non-candidate tripwire (R4) test
these assumptions; they do not prove C0/C1 on the full input. Not covered:
C2--C4; residual couplings below $\kappa$ or that evade pads and the audit;
limitation F6 (coupling outside the closed family that still looks locally
consistent).

\paragraph{Audit bound.}
Let $A$ be the stated audit population and fix the affected records
independently of the audit draw. Under (I), sample-only stock is a valid
oracle. If at least a fraction $\kappa$ of $A$ differs, a uniform sample
of $s\le |A|$ records without replacement misses them with probability
at most $(1-\kappa)^s$. The conservative with-replacement sizing rule is
\begin{equation}
s=\bigl\lceil\ln\varepsilon/\ln(1-\kappa)\bigr\rceil.
\label{eq:audit-s}
\end{equation}
For $\varepsilon=0.05$, $\kappa=0.01$, $s=299$; if $|A|<s$, audit all of $A$.
The bound requires that the stated audit executes with a valid oracle.
The probability is over a uniform draw under a non-adaptive affected set
(independent of the sample). The shipped CLI defaults to a fixed audit seed
(20260927), so a repeated update with an unchanged record order redraws the
same sample; that seed is recorded for reproducibility and does not give
independent risk reduction across releases, nor protection against faults
chosen after the sample is known.
For the historical bounded timings, $A$ contains only candidate-bearing
records; the final all-record policy uses $A=R$, with timing in the locked campaign.
In the historical adversarial evaluation, 3 of 400 trials shipped when
at least 1\% of its audit population was affected.
\begin{plainbox}{What the audit does \emph{not} cover}
It is not a 5\% per\nobreakdash-line error rate, nor a proof that the program is entry\nobreakdash-local, nor a bound on shipping
\textit{any} incorrect output.
For one affected record among 5{,}000, a uniform 299\nobreakdash-record audit alone misses it with probability
$\approx(1-1/5000)^{299}\approx 0.94$.
Pads may catch some such faults; we establish no general bound for faults below $\kappa$.
Structured faults that evade the pads and the sample remain possible (limitation F6).
C2--C4 are refused when observed (R1--R6), not ``covered'' by $s{=}299$.
Assumption~(I) is required for the sample\nobreakdash-only stock oracle; it is not proved by running the audit.
\end{plainbox}

\paragraph{Cost model.}
With stock time $T_{\mathrm{stock}}$,
\begin{equation}
T_{\sys}\approx a\,f_\Delta\,T_{\mathrm{stock}}+c\,T_{\mathrm{stock}},
\label{eq:cost}
\end{equation}
where $a$ scales selection and $c$ collects replay, check, audit, and parsing.
Fallback when $f_\Delta>(1-c)/a$; default threshold $0.80$ with no timing history.
Programs whose rows factor as $f(r,e,o,g(E))$ and satisfy the exact-merge
assumptions fit; C2--C4 remain refuse
classes. \sys never reads the program's source at run time.

\section{Design}
\label{sec:design}

\begin{figure*}[t]
\centering
\resizebox{\textwidth}{!}{%
\begin{tikzpicture}[
  font=\scriptsize,
  stage/.style={draw=cTare!80!black, rounded corners=2pt, fill=cTare!7, minimum height=0.95cm, text width=2.45cm, align=center, inner sep=2pt},
  gate/.style={draw=cGrey, rounded corners=2pt, fill=gray!10, minimum height=0.95cm, text width=2.45cm, align=center, inner sep=2pt},
  store/.style={draw=gray, fill=gray!8, cylinder, shape border rotate=90, aspect=0.18, minimum height=0.85cm, minimum width=1.55cm, align=center, inner sep=1pt},
  io/.style={draw, rounded corners=2pt, minimum height=0.7cm, text width=2.2cm, align=center, inner sep=2pt},
  arr/.style={-{Stealth[length=1.8mm]}, thick, cTare!80!black},
  rarr/.style={-{Stealth[length=1.8mm]}, thick, cBad},
  lab/.style={font=\tiny, fill=white, inner sep=1pt},
  node distance=0.5cm and 0.62cm]
\node[stage] (s1) at (0,0) {\textbf{1. Entry diff}\\ split $E'$ into stable $S$, changed $\Delta$};
\node[stage, right=of s1] (s2) {\textbf{2. Contract inference}\\ probe halves, fit each column};
\node[gate, right=of s2] (s3) {\textbf{3. Coupling gate}\\ pin, side files, groups};
\node[io, fill=cBad!8, draw=cBad, right=0.9cm of s3] (refuse) {\textbf{Refuse}\\ fail + dump stock};
\node[stage, below=1.25cm of s3] (s4) {\textbf{4. Selection}\\ all records vs.\ $\Delta$ $\rightarrow$ $C_r$};
\node[stage, left=of s4] (s5) {\textbf{5. Batched replay}\\ batches vs.\ their candidates};
\node[gate, left=of s5] (s6) {\textbf{6. Check or refuse}\\ pads, then a sampled audit};
\node[io, fill=cTare!12, draw=cTare, left=0.75cm of s6] (ship) {\textbf{Ship}\\ under contract + audit};
\node[store, above=0.75cm of s1] (cache) {cache\\ $P(R,E)$\\[-1pt]{\tiny (also read by 4, 5)}};
\node[store, above=0.75cm of s2] (contract) {contract\\ (per argv)};
\node[io, fill=gray!8, left=0.75cm of s1, text width=1.6cm] (in) {records $R$\\ old $E$, new $E'$};
\draw[arr] (in) -- (s1);
\draw[arr] (s1) -- node[lab, above]{$S$, $\Delta$} (s2);
\draw[arr] (s2) -- node[lab, above]{contract} (s3);
\draw[arr] (s3) -- node[lab, left]{authorize, pin $\pi$} (s4);
\draw[arr] (s4) -- node[lab, above]{$C_r$} (s5);
\draw[arr] (s5) -- node[lab, above]{replay body} (s6);
\draw[arr] (s6) -- node[lab, above]{pass} (ship);
\draw[arr, densely dashed] (contract) -- (s2.north);
\draw[arr, densely dashed] (cache) -- (s1.north);
\draw[rarr] (s3) -- node[lab, above]{R1--R3} (refuse);
\coordinate (bus) at ($(s5.south)+(0,-0.5)$);
\draw[thick, cBad] (s5.south) -- node[lab, right]{R4} (s5.south |- bus);
\draw[thick, cBad] (s6.south) -- node[lab, right]{R5, R6} (s6.south |- bus);
\draw[rarr] (s6.south |- bus) -- (refuse.south |- bus) -- (refuse.south);
\node[anchor=north west, font=\tiny, text=cBad!85!black, text width=15.2cm, align=left] at ($(s6.south west |- bus)+(-0.1,-0.08)$)
  {\textbf{Refusal rules} (each exits with failure and dumps stock when the tool runs): R1 no contract;\; R2 size-dependent, no pin;\; R3 in-situ
   check fails;\; R4 a non-candidate pair appeared;\; R5 a pad changed a line;\; R6 the audit differs.};
\begin{scope}[on background layer]
\node[fill=cTare!3, rounded corners=4pt, fit=(s1)(s2)(s3)(cache)(contract), inner sep=5pt, label={[font=\tiny\bfseries, text=cTare!70!black, anchor=south west]north west:DECIDE: is exact replay possible, and how?}] {};
\node[fill=cTare!3, rounded corners=4pt, fit=(s4)(s5)(s6), inner sep=5pt, label={[font=\tiny\bfseries, text=cTare!70!black, anchor=south west]north west:REPLAY AND CHECK}] {};
\end{scope}
\end{tikzpicture}%
}

\caption{\textbf{How \sys updates an annotation.} Blue boxes do work; gray boxes are decision points. The top row
decides whether the program's coupling can be neutralized; the bottom row reruns the unmodified program on candidates
only and checks the result. Dashed arrows read stored state. Every refusal (red) exits with failure status and dumps a
full stock body when the tool still runs (stock failure is reported beside the refuse reason). A shipped replay is exact
only under the contract; the audit is a separate prevalence bound. Finite checks do not prove the contract.}
\label{fig:pipeline}
\end{figure*}
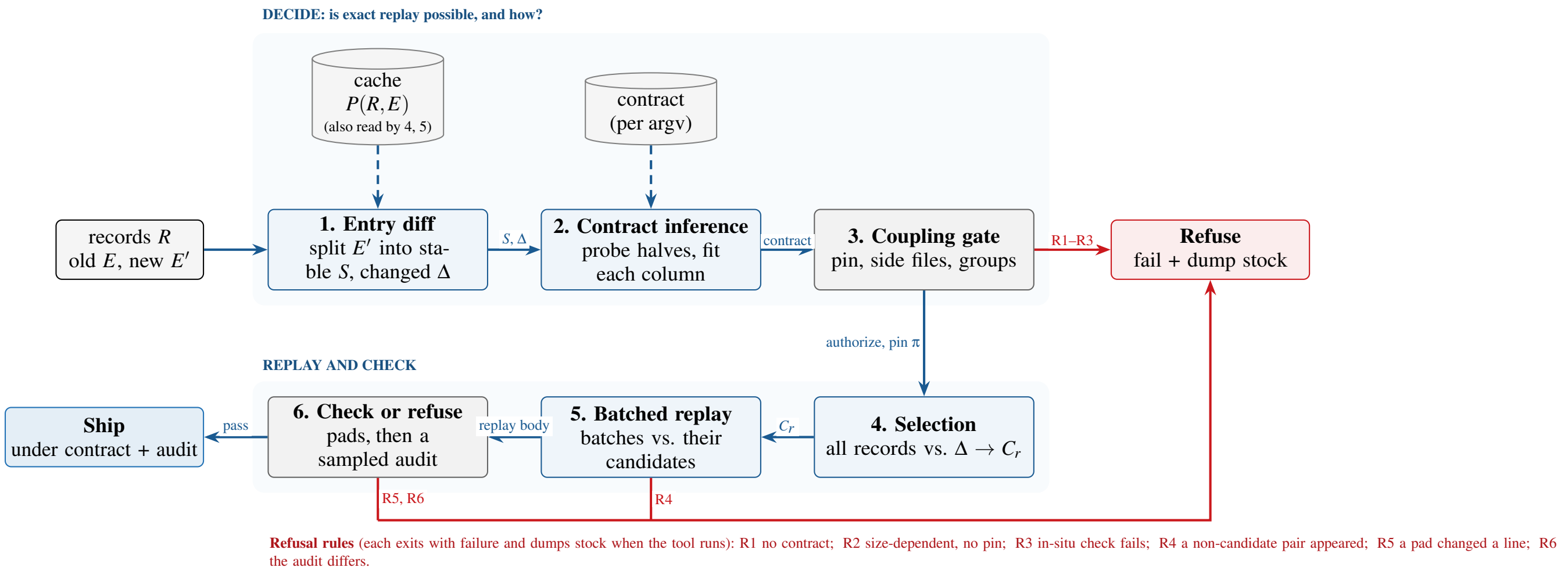

\sys updates a cached annotation in six stages (\cref{fig:pipeline}). The first three decide \textit{whether} exact
replay is possible and \textit{how}; the last three perform it and check it. Any failed decision or check refuses
with failure status and dumps a full stock body when the tool still runs. Passing these finite checks supports the
contract on that run; it does not prove it.

\subsection{Stage 1: Entry Diff}

\sys splits both releases into entries and computes a \textit{strict hash} of each: SHA-256 over the entry's full
text, including its name, accession, description, length, Pfam's curated thresholds (GA, TC, NC), and every model line,
after dropping only lines tagged DATE (build date), BM and SM (the commands that built and calibrated the model), or COM
(comments). A renamed or re\nobreakdash-thresholded family therefore counts as changed. Entries of the new release whose hash appears in the old one are stable; all
others are changed. A cache keyed on the whole database file would miss on every entry after a release; a cache keyed
on entry hashes keeps every stable entry, 86.6\% of Pfam for 38.1$\rightarrow$38.2.

\subsection{Stage 2: Contract Inference}
\label{sec:contract}

Without reading the program's code, \sys probes a sample of records three times---whole reference and each half---and
aligns lines by record and entry. If every whole\nobreakdash-run line appears in exactly one half, the output \textit{decomposes}
over entries; a whole\nobreakdash-only line means unreplayable interaction (refuse). A half\nobreakdash-only line is allowed only when the
user declares competing entries (\cref{sec:gate}). For each numeric column it tests identity or scaling with entry
count, total length, or hits per record; a fit within printed precision is required, else refuse. Ranks and labels are
left to the program. The resulting \textit{contract} is computed once per command and reused across releases.

\subsection{Stage 3: The Coupling Gate}
\label{sec:gate}

The gate turns the contract into a decision. Users may add declarations, which are data, never program names.

\textit{Size dependence $\rightarrow$ pin.} If a column scales with the reference size, \sys needs a program option that
sets that size. Users declare it (\texttt{-{}-pin~'-Z~\{entry\_count\}'}), and every experiment in this paper used a
declared pin. \sys can also search for one, by reading the program's help text and trying each numeric option until one
moves the column by exactly the predicted ratio; we tried this search only on HMMER and DIAMOND, and we have not
evaluated its risks, such as an option like a random seed moving a column by coincidence. With the option set to the
new release's value, the program prints on a subset the numbers it would print on the whole.

\textit{No size dependence $\rightarrow$ no pin.} If every column fits identity, \sys replays without a pin. This is the
case for Infernal under Rfam's recommended command, which derives its search size from the query sequence instead.

\textit{Side files and competing entries.} Rfam's recommended command passes a ``clan'' file listing families that
compete. On a subset, \sys rewrites that file to the families present, because the program rejects names it cannot
find. If any member of a declared group changes, \sys re\nobreakdash-searches the whole group, because a change to one member
can reveal a hit of another that the cache never recorded.

The gate applies R1--R6 from \cref{sec:model} (R1--R3 on the probe; R4--R6 on the update). Every refusal exits with
failure status and dumps a full stock body when the tool still runs.
\dec{refuse\_pin} (R2/R3) differs from \dec{refuse} only in reporting that a working pin could enable replay.
Before replaying, the gate also refuses when $f_\Delta$ exceeds the break\nobreakdash-even point $(1-c)/a$ of \cref{eq:cost}
(default 0.80 with no timing history).

\subsection{Stage 4: Selection}

Changed entries may produce hits the cache has never seen. \sys runs the program on all records against the changed
entries $\Delta$ alone, with the pin set. A record's \textit{candidates} are its cached hits on stable entries plus its
new hits:
\begin{equation}
C_r = \{e \in S : (r,e) \in P(R,E)\} \;\cup\; \{e : (r,e) \in P_\pi(R,\Delta)\}.
\label{eq:cand}
\end{equation}
Selection compares every record with every changed entry, so its cost is proportional to the changed fraction of the
reference. It is the one cost \sys cannot avoid (\cref{sec:cost}).
Candidate completeness for the merge is
$\mathrm{Hits}(r,E_{\mathrm{new}})\subseteq\mathrm{Cached}(r)\cup\mathrm{Discovered}(r)$, which holds here because $Z$ is pinned and
reporting thresholds are per family (\cutga); without those conditions an unchanged pair that was previously below a
global threshold can become reportable and be missing from both sets.

\subsection{Stage 5: Batched Exact Replay}
\label{sec:replay}

Cached lines cannot be returned as they are: their numbers used the old release. \sys reruns the program on the
candidates with the pin set so the program reprints. A naive all\nobreakdash-candidates universe wastes work---on one genome the
union covers 3{,}719 families while the mean protein has two---so proteins are batched against only their own
candidates and stitched in record order, choosing the lowest predicted\nobreakdash-cost plan among a searched family:
proteins sorted by candidate rarity, then equal contiguous chunk sizes on a grid. That family is not a global
partition optimizer. Predicted cost is
\begin{equation}
T_{\text{replay}} = \alpha k + \beta \sum_b |U_b| + \gamma \sum_b |B_b|\,|U_b|,
\label{eq:batch}
\end{equation}
with $k$ batches, $|B_b|$/$|U_b|$ proteins/families in batch $b$, and $\alpha,\beta,\gamma$ start, prepare, and
compare costs. HMMER's per\nobreakdash-protein blocks depend only on that protein's hits. Tripwire R4: a pair outside
$C_r$ refuses.

\subsection{Stage 6: Check or Refuse}
\label{sec:check}

The check falsifies C0/C1 on the run: replay with \textit{pads} (left\nobreakdash-out families added back). The
\textit{neighbor pad} adds file\nobreakdash-order neighbors of every candidate (R5; catches C2 as in DIAMOND,
\cref{sec:eval-safety}). \textit{Random pads} add three random left\nobreakdash-out samples drawn from the non\nobreakdash-candidate
union, each of width $\min(\max(200,|U|),|\mathrm{non}|)$ where $|U|$ is the candidate\nobreakdash-union size,
on $s{=}50$ proteins.
If a dependency affects $k$ of $n$ proteins and a pad hits the responsible family with probability $q$,
\begin{equation}
P_{\text{miss}} \approx h + (1-h)(1-q)^3, \qquad h = \binom{n-k}{s}\Big/\binom{n}{s}.
\label{eq:miss}
\end{equation}
Pads test C0/C1; the audit bound is R6 with $s$ from \cref{eq:audit-s}.
Three audit populations appear in this paper and must not be conflated (\cref{tab:config}).
The historical Bernoulli policy ($p{=}0.02$, floor 20) and the historical bounded policy ($s{=}299$ on records with at
least one replay candidate) each keep their own freeze and timing; the $\HeadlineBoundedMedian\times$ figure is for the
candidate\nobreakdash-bearing population only.
The \emph{final} shipped policy for the headline is the locked all\nobreakdash-record campaign: $s{=}299$ drawn from
\textit{all} query records, including empty candidate sets, timed at $\HeadlineMedian\times$
(\cref{tab:config,tab:allrecord-genomes}).
The audit oracle reruns stock on the sampled records alone.
Record\nobreakdash-subset invariance of that oracle (sample\nobreakdash-only stock equals the projection of full\nobreakdash-query stock) is
assumption~(I) of \cref{prop:exact}; it is not separately proved by the campaign.
An optional first\nobreakdash-use mode compares the first two updates to a full run.

\subsection{Cost Model}
\label{sec:cost}

\cref{eq:cost} in \cref{sec:model} is the working model: selection scales with $f_\Delta$, and $c$ collects replay,
check, audit, and parsing. For $a{\ge}1$ and $c{\ge}0$, the speedup $1/(a f_\Delta + c)$ never exceeds $1/f_\Delta$
or $1/c$; without that unit\nobreakdash-cost selection assumption the immediate ceilings are $1/(a f_\Delta)$ and $1/c$.
The first run also fills the cache (a full run plus contract probes); with fill time $F$ and update time $U$,
\sys becomes cheaper than repeated full runs after $n \ge (F - T_{\mathrm{stock}})/(T_{\mathrm{stock}} - U)$ releases.

\section{Implementation}
\label{sec:impl}

\sys's core is 12{,}555 lines of Python, with 358 test functions. Its core contains no program names: small parsers read
profile databases, covariance\nobreakdash-model databases, FASTA files, and tabular outputs, and users declare pins, side files,
groups, and reporting thresholds as data. The cache is a SQLite file keyed by entry hash and record. A typical
invocation wraps the normal command shown below. The cache is namespaced by a hash of the command line with the
reference path removed, so changing any option, including the thread count, starts a new cache and a new contract.
This is conservative: repeated stock runs were byte\nobreakdash-identical in every timed experiment. The
namespace also includes a SHA-256 of the program binary, so upgrading a program in place starts a new cache.
\begin{lstlisting}
acts run --strategy reference --input records.faa \
  --reference pfam38.2.hmm --prep 'hmmpress {reference}' \
  --pin '-Z {entry_count}' --cache cache.sqlite -- \
  hmmscan --cut_ga --noali --cpu 4 \
  --tblout {output:tblout} --domtblout {output:domtblout} \
  {reference} {input}
\end{lstlisting}
The installed CLI is \texttt{acts} (\texttt{python -m acts}); placeholders are caller paths.
\texttt{-{}-input} is required for the reference strategy; \texttt{-{}-prep} is caller\nobreakdash-declared
(sidecar builds are not inferred).

\section{Evaluation}
\label{sec:eval}

We ask four questions. \textbf{Q1:} Is \sys's output identical to a full run? \textbf{Q2:} How much faster is an
update, and does the cost model predict it? \textbf{Q3:} Does \sys refuse when replay would be wrong? \textbf{Q4:} How
does it compare with existing tools?

\textbf{Method.} Every experiment followed the same protocol, designed so that a failure means the method failed rather
than that the test was blind. We first read the program's source for behaviors that could make a comparison fail for
unrelated reasons, then ran a pilot on separate data, then committed the design and pass/fail rules to version control
before the measured run. Each test included a case that must pass and a case that must fail, and a minimum number of
output lines below which a result counts as inconclusive. Measured runs used genomes drawn by a fixed random seed;
new draws were disjoint from their corresponding pilots. The all\nobreakdash-record campaign reuses the historical nine\nobreakdash-genome
timing cohort for retiming under the locked policy. One rule changed after its pre\nobreakdash-registration, during a measured run; we disclosed the change and
reran every control (\cref{sec:eval-safety}).

\textbf{Setup.} HMMER 3.4 (\texttt{hmmscan}) with Pfam 38.1 and 38.2; Infernal 1.1.5 (\texttt{cmscan}) with Rfam 15.0
and 15.1; DIAMOND 2.2.5 and BLAST+ 2.14.1 with Swiss\nobreakdash-Prot 2026\_01 and 2026\_03; queries are bacterial genomes from NCBI
RefSeq. Server timings use exclusive AMD EPYC 7542 compute with shared Lustre scratch for inputs and caches;
SQLite state restoration sits outside the update timer; an unscored stock warm\nobreakdash-up precedes timed arms;
a wall\nobreakdash-quality flag excludes a genome when its max/min repeat ratio exceeds 1.5.
Laptop timings use an Apple laptop shared with other work and are indicative only.

\subsection{Metrics}
\label{sec:metrics}

\textbf{Exactness and assurance.} We separate three statements and never equate them:
(i)~\emph{exact under the contract}---\cref{prop:exact} under C0/C1 and (D)--(O), (C), and (I);
(ii)~\emph{body\nobreakdash-line agreement on evaluated runs}---a run is \dec{exact} in tables if the ordered
sequence of nonempty, noncomment body\nobreakdash-line contents (including within\nobreakdash-line bytes) equals the stock
run's; blank and comment lines are dropped, and LF/CRLF are equated by line splitting;
(iii)~\emph{probabilistic validation}---pads plus a sample that bounds missing faults above a stated prevalence in a
stated population, not the probability of shipping any incorrect output.
Lines within a run share a pin and a reference, so they are not independent observations.
We report compared-line counts and observed differences without inferring a per\nobreakdash-line error probability.
\textbf{Speedup.} For genome $g$ with $r$ alternating runs per arm, $S_g$ is the median stock wall time divided by the
median \sys wall time, both measured around the whole process. Across genomes we report the median of $S_g$ with a 95\%
percentile bootstrap interval, the geometric mean, and the range. Differing lines are classified as whitespace\nobreakdash-only,
value\nobreakdash-only, missing, extra, or reordered. Safety outcomes are \dec{exact}, \dec{safe-refuse}, or \dec{unsafe}
(any \dec{unsafe} fails the claim). Wall time $T$ is monotonic elapsed time of one complete invocation excluding
\texttt{hmmpress}. Baseline overhead uses paired $\rho = T^{\text{tool}}/T^{\text{stock}}$ with a 95\% $t$-interval.

\subsection{Q1: Is the Output Exact?}
\label{sec:eval-exact}

\begin{table}[t]
\caption{Exactness experiments. ``Lines'': output lines compared byte for byte with a stock run on the new release
(HMMER: hit plus domain table). ``Differ'': lines that differed. ``Unpinned'': the same replay without the pin.
``Replay'' runs the replay engine alone; ``tool'' runs \sys end to end, with batching and the check.}
\label{tab:exact}
\centering\footnotesize
\setlength{\tabcolsep}{3pt}
\begin{tabular}{@{}lrcr@{}}
\toprule
Experiment & Lines & Differ & Unpinned differ \\
\midrule
HMMER replay, 300 proteins & 1{,}079 & 0 & 534 of 534 \\
HMMER replay, 3 genomes & 51{,}406 & 0 & 26{,}094 of 26{,}127 \\
HMMER tool, 2 genomes (laptop) & 35{,}434 & 0 & not run \\
HMMER tool, 9 genomes (server) & 155{,}577 & 0 & not run \\
HMMER tool, 48 diverse genomes & 463{,}225 & 0 & not run \\
Infernal replay, 10 genomes & 4{,}217 & 0 & no pin needed \\
Infernal tool, 1 family removed & 459 & 0 & no pin needed \\
Infernal tool, 10 new genomes & 4{,}307 & 0 & no pin needed \\
\midrule
Total & 715{,}704 & 0 & \\
\bottomrule
\end{tabular}
\end{table}

Across the eight experiments in \cref{tab:exact}, not one of 715{,}704 compared lines differed from the stock run.
This is observed agreement on those runs, not a probability guarantee for other inputs.
Without the pin, nearly every line differed: 26{,}094 of 26{,}127 hit lines on three genomes ($d = 99.87\%$)
and all 534 on 300 proteins. In these experiments, the pin was necessary; with \cutga, pinned replay matched stock.

\textbf{Diverse genomes and needless refusals.} On 53 further genomes (14 archaea, 30 bacteria, 2 eukaryotes, and 7
metagenome\nobreakdash-assembled genomes), \sys shipped 48 updates, all byte\nobreakdash-identical, and produced no unsafe result. It
refused the other 5 although their replays matched stock: 3 because two domain\nobreakdash-table rows of one protein shared the key
(protein, family, E-value), and 2 because a text column held some values that parse as numbers. The \textit{Drosophila}
proteome, with 96 proteins of more than 20 domains, was one of the latter; three further eukaryotes did not finish within
12 hours and one bacterial job crashed on a storage error, so neither is counted. The duplicate\nobreakdash-key rule also refused
one of the ten server genomes below. These refusals cost speed, not correctness. After the measured runs we fixed both
rules: repeated keys are told apart by their order, and a column that is mostly text is passed through. All six
needlessly refused HMMER genomes, including the server genome and \textit{Drosophila}, then shipped byte\nobreakdash-identical
updates (a follow\nobreakdash-up, not part of any locked draw).

\textbf{Where exactness ends.} When HMMER reports hits by E-value threshold instead of \cutga, the replay matched on 300
proteins but failed on 1{,}000, differing on 3 hit lines and 5 domain lines. Hits near the threshold move in or out of a
protein's list, which changes that protein's statistics; this mode is not entry\nobreakdash-local and we do not claim it. The early
pass on 300 proteins is why we require a minimum number of lines before accepting any result.

\textbf{A second program.} Infernal, which searches RNA families, behaved as its source code predicted: under Rfam's
recommended command every numeric column fits identity, so no pin is needed. Replaying each genome's hits was
byte\nobreakdash-identical on ten genomes (4{,}217 lines) with and without clan competition. In that command, overlapping hits of
competing families are marked rather than deleted, so removing a family cannot reveal a hidden hit; removing the family
\texttt{GlmZ\_SraJ} from one genome's reference was exact through the shipped tool (459 lines). With the non\nobreakdash-default
option \texttt{-{}-oskip}, which deletes marked overlaps, the same removal was refused, because fourteen overlap marks depended
on families that had no remaining line. Through the shipped tool on ten further genomes (Rfam 15.0$\rightarrow$15.1, three alternating runs per arm), every one of the
60 updates was byte\nobreakdash-identical to its paired stock run in both variants, with 4{,}307 body lines per variant against a
pre\nobreakdash-registered floor of 2{,}000.

\subsection{Q2: How Much Faster?}
\label{sec:eval-time}

\begin{figure}[t]
\centering
\includegraphics[width=\linewidth]{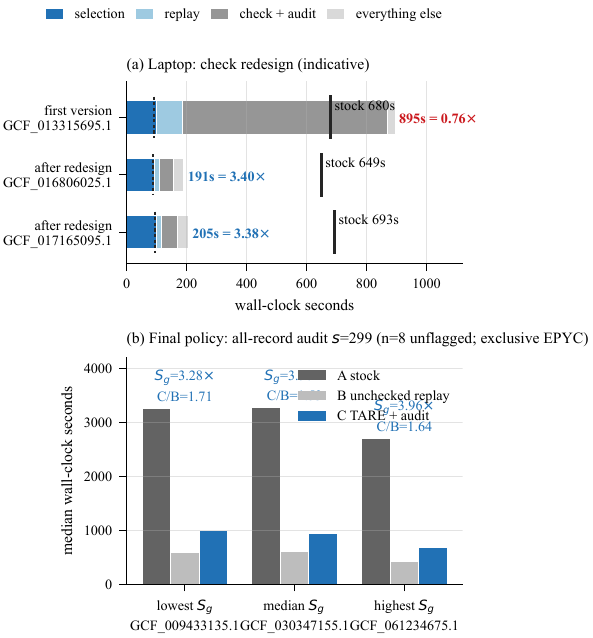}
\caption{\textbf{Where the time goes, then what the final policy costs.}
(a)~Laptop (indicative): stacked phases; black tick $=$ stock.
The first version spent most of its time in the check and was slower than stock; after redesign, two genomes beat stock.
(b)~Final all\nobreakdash-record policy on exclusive EPYC ($s{=}299$, unflagged genomes): median walls for stock (A),
unchecked pinned replay (B), and \sys with audit (C). Annotated $S_g{=}A/C$ and certificate cost $C/B$.
This panel is the locked campaign, not the historical Bernoulli bars.}
\label{fig:phases}
\end{figure}

\Cref{fig:phases} shows that the replay and check designs decide whether \sys is faster than a full run. Our first
version replayed every protein against all candidates and checked the result with four more full\nobreakdash-size replays. On the laptop, with three alternating runs per arm, it took a median of 895\,s against a median stock time of 680\,s
($S_g = 0.76\times$; CV 3.7\% for \sys and 1.7\% for stock); the check alone took a median of 668\,s. Batching made the
replay $5.2\times$ cheaper and sampling made the check 17.9--19.3$\times$ cheaper (ratios of the first version's median check time to each shipped run's). On two fresh genomes of 4{,}583 and 5{,}019
proteins, \sys replayed in 107 and 111 batches and took 191\,s and 205\,s against stock runs of 649\,s and 693\,s: $S_g = 3.40\times$ and $3.38\times$
(geometric mean $3.39\times$), with byte\nobreakdash-identical output. These are single runs on a shared machine, so we report no
confidence interval for them; the server headline repeats each arm three times.

\begin{figure}[t]
\centering
\includegraphics[width=\linewidth]{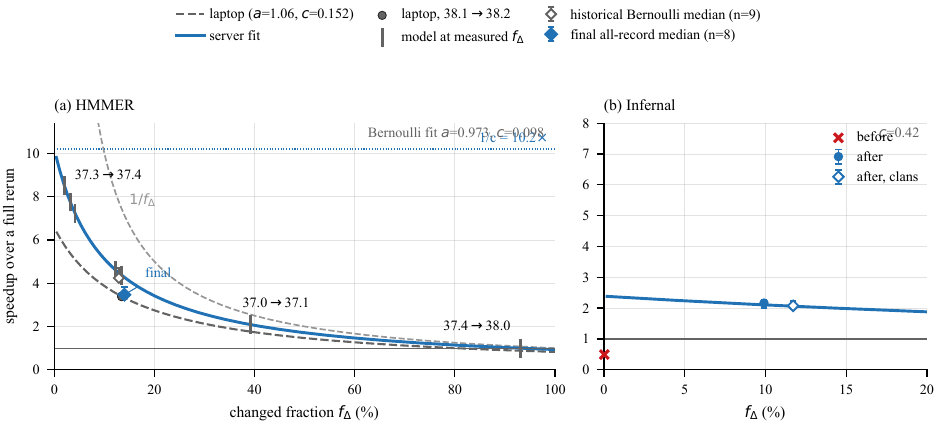}
\caption{\textbf{The cost model across release sizes.} Speedup $1/(a f_\Delta + c)$.
(a)~HMMER: server fit from the nine historical Bernoulli genomes ($a{=}0.973$, $c{=}0.098$, solid) and laptop fit
(dashed). Open gray diamond: historical Bernoulli median at $f_\Delta{=}13.4\%$ (n$=$9). Filled blue diamond:
\textbf{final} all\nobreakdash-record median (n$=$8 unflagged) with 95\% CI---the paper's headline, not the fit's
training policy. Ticks: model at other measured Pfam $f_\Delta$; gray dashed $1/f_\Delta$; dotted $1/c$.
(b)~Infernal: first version (cross) and shipped tool (medians of ten genomes, 95\% bootstrap CIs).}
\label{fig:model}
\end{figure}

\textbf{The cost model.} With $f_\Delta = 0.134$, the selection term predicts 87\,s and 93\,s on those genomes; selection measured 91\,s and 100\,s, a
relative error of 4.3\% and 7.1\%. The remaining costs, as a fraction of each stock run, average $c = 0.152$ (arithmetic mean of 0.153 and 0.151), so even a release that changed
nothing could be at most $6.6\times$ faster with today's overheads (\cref{fig:model}); the check, audit, and parsing are
therefore the next targets. On the server, the same fit over the nine headline genomes gives $a = 0.973$ and
$c = 0.098$, a ceiling of $10.2\times$, and at headline $f_\Delta{=}0.134$ predicts $4.37\times$ versus the measured
bounded\nobreakdash-audit median $\HeadlineBoundedMedian\times$. The fit uses the lighter Bernoulli policy; it does not
isolate the causes of the gap to the bounded policy ($s{=}299$).
The laptop cache fill took 871\,s and 852\,s (single runs),
1.2--1.3$\times$ a stock run, so \sys pays for itself from the first update (projected from these indicative times).

\begin{figure*}[t]
\centering
\includegraphics[width=\textwidth]{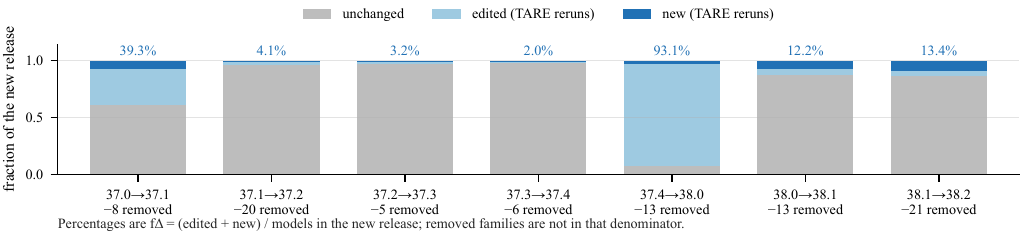}
\caption{\textbf{Edits, not just new families, decide how much discovery searches.} Strict hashes of seven consecutive
Pfam releases. Gray is unchanged. Light blue is edited and dark blue is new---the changed set $\Delta$. The label is
$f_\Delta = (\text{edited} + \text{new}) / Z_{\text{new}}$. Pfam 37.4$\rightarrow$38.0 has $f_\Delta{=}93.1\%$ of the new release: 22{,}965 edited and 822 new, with 1{,}758 unchanged, of 25{,}545.
Pfam 38.1$\rightarrow$38.2 has $f_\Delta{=}13.4\%$: 1{,}378 edited and 2{,}674 new, with 26{,}082 unchanged, of 30{,}134, the same pair as
\cref{fig:churn}. Replay candidates and pads/audit add work beyond $\Delta$. The smallest change is 37.3$\rightarrow$37.4, at 2.0\%.}
\label{fig:pairs}
\end{figure*}

\textbf{Other releases.} Strict hashes of all seven consecutive Pfam pairs (\cref{fig:pairs}) put the changed fraction
between 2.0\% and 93.1\%. At those measured values the server model predicts $4.6$--$8.5\times$ for the four other pairs that
changed at most 12.2\%, $2.1\times$ for 37.0$\rightarrow$37.1 (39.3\%), and $1.00\times$ for 37.4$\rightarrow$38.0, which rewrote
93.1\% of Pfam's models. That release was a rebuild, not churn: 22{,}925 of its 22{,}965 edited models carry a new
accession version, which \texttt{hmmscan} prints, and 19{,}976 have different model parameters. For such a release replay
cannot pay, so the gate falls back to a full run (\cref{sec:gate}): with the server fit the break\nobreakdash-even fraction is
0.93, and on a 40\nobreakdash-model subset of this release ($f_\Delta = 0.975$) the default threshold fell back and the output
matched stock.
The headline times the 38.1$\rightarrow$38.2 pair. On the measured release chain (three fresh genomes, seed 20261222,
exclusive EPYC 7542, $r{=}3$ per step), a cache filled at Pfam 37.3 and updated through 37.4, 38.0, 38.1, and 38.2:
two of three genomes shipped exact (byte\nobreakdash-identical) at every step. Per\nobreakdash-step $S_g$ was
7.08$\times$/8.92$\times$ at 37.4, 0.91$\times$/0.68$\times$ at 38.0, 4.29$\times$/4.54$\times$ at 38.1, and
4.01$\times$/5.35$\times$ at 38.2. Cumulative wall including the fill, $C_{\mathrm{stock}}/C_{\sys}$ after the last
step, was 1.66 and 1.71 (after 37.4 alone: 1.31 and 1.46). The 38.0 step is the case the break\nobreakdash-even fallback
is for: this run used frozen code from before that fallback, so it shows the cost without it. The third genome was
refused at the fill (37.3) for a duplicate row key, fell back to a full scan, and shipped no chained output---the same
needless\nobreakdash-refusal class a later row\nobreakdash-key fix addressed; the measured chain used the pre\nobreakdash-fix
code and counts a scored FAIL.

\textbf{Server headline (all\nobreakdash-record).} On a dedicated node, across the same nine genomes with three alternating
stock and \sys runs each under the locked all\nobreakdash-record bounded audit ($s{=}299$), the median of $S_g$ over
\HeadlineN\ non\nobreakdash-FLAGGED genomes is $\HeadlineMedian\times$ (95\% genome\nobreakdash-cluster bootstrap CI
\HeadlineCIlow--\HeadlineCIhigh; geometric mean $\HeadlineGeoMean\times$; range \HeadlineRangeLow--\HeadlineRangeHigh$\times$).
\FlaggedCount\ genome was wall\nobreakdash-quality FLAGGED and excluded from that median (\FlaggedList; all\nobreakdash-genome
sensitivity $\SensitivityAllGenomes\times$). Every scored cell was body\nobreakdash-equal to stock
(\CampaignBodyLines\ compared tblout+domtblout lines). Cost of checking versus unchecked planner replay is
$\CostOfCheckingMedian\times$ (C/B; range \CostOfCheckingRangeLow--\CostOfCheckingRangeHigh$\times$ on unflagged genomes).

\textbf{Historical candidate\nobreakdash-population timings.} On the same nine genomes, the earlier candidate\nobreakdash-population
bounded audit ($s{=}299$) gave median $S_g{=}\HeadlineBoundedMedian\times$ (95\% CI \HeadlineBoundedCIlow--\HeadlineBoundedCIhigh),
and Bernoulli ($p{=}0.02$, floor 20) gave $4.24\times$ (CI 4.17--4.68); cache fill was a median $1.37\times$ stock under the
bounded policy. One pre\nobreakdash-registered genome refused at fill and three 4\nobreakdash-hour timeouts were rerun at 8 hours
(recorded before rerun).

\textbf{Infernal, with a reduced check.}
\begin{plainbox}{Infernal reduced check (long DNA)}
When one record holds $>$half the input residues, the sampled audit and the random pads are skipped (an audit of the whole genome \textit{is} a stock run).
What remains: the neighbor pad and the in\nobreakdash-situ pin check. Speedups below must be read with that caveat; every known\nobreakdash-bad control still refused.
\end{plainbox}
The need for that rule showed up on a single\nobreakdash-family removal: only 457 of Rfam 15.1's 4{,}227 families (10.8\%) changed, yet the update took 666\,s against a 328\,s stock run ($0.49\times$), exact but slower, because a bacterial genome is one DNA sequence (or a handful) and the full audit covered everything. Budgeting audit and random pads by residues as above leaves HMMER's decisions and proteome output unchanged.

With this change, Infernal updates on ten fresh genomes ran a median $S_g = 2.16\times$ faster than stock without clan
competition (95\% bootstrap CI 2.00--2.28; geometric mean $2.07\times$; range 1.41--2.37) and $2.07\times$ with it (CI
1.97--2.23; geometric mean $2.10\times$; range 1.96--2.33), against cost\nobreakdash-model predictions of $2.08\times$ and $2.05\times$.
These are laptop timings with three alternating runs per arm: the largest per\nobreakdash-arm CV was 29\% (flagged, not dropped, as
pre\nobreakdash-registered), and the largest per\nobreakdash-genome prediction error was 32\%.

\textbf{Infernal payback is slow.} Cache fill was a median $4.45\times$ stock (payback $\approx$6.4 updates); Infernal is worthwhile today mainly across several Rfam releases.

\subsection{Q3: Does \sys Refuse When It Should?}
\label{sec:eval-safety}

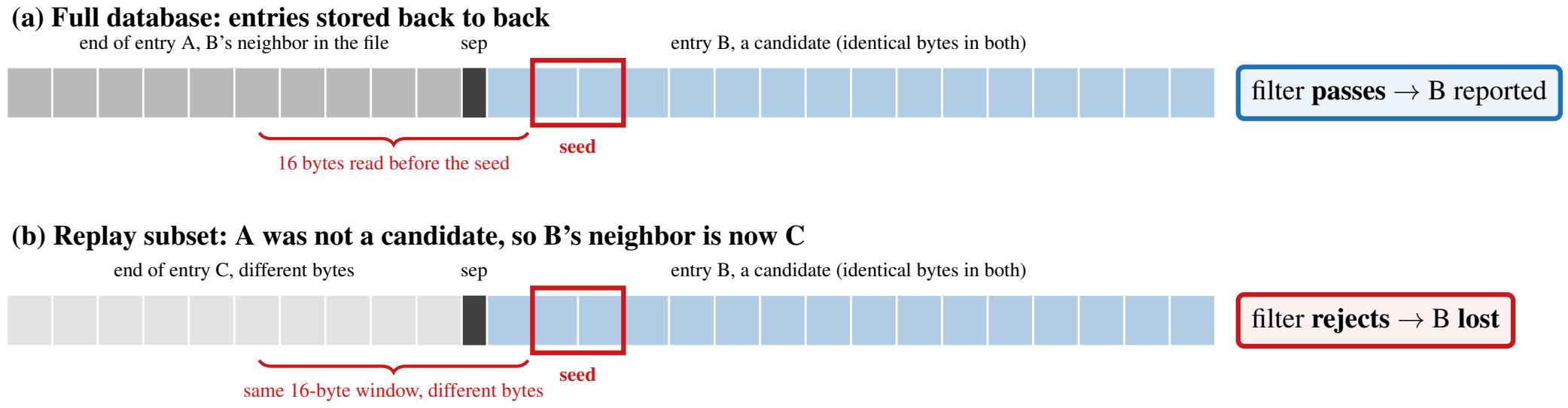
\begin{figure*}[t]
\centering
\begin{tikzpicture}[font=\scriptsize, x=0.40cm, y=0.5cm]
  \def\cellA{gray!55}\def\cellB{cTare!35}\def\cellC{gray!22}
  \newcommand{\strip}[3]{
    \foreach \i in {0,...,9} \draw[fill=#2, draw=white, line width=0.5pt] (\i,#1) rectangle ++(1,0.9);
    \draw[fill=black!75, draw=white] (10,#1) rectangle ++(0.55,0.9);
    \foreach \i in {10.55,11.55,...,25.55} \draw[fill=\cellB, draw=white, line width=0.5pt] (\i,#1) rectangle ++(1,0.9);
    \node[font=\tiny, above] at (5,#1+0.9) {#3};
    \node[font=\tiny, above] at (10.27,#1+0.9) {sep};
    \node[font=\tiny, above] at (18.5,#1+0.9) {entry B, a candidate (identical bytes in both)};
    \draw[very thick, cBad] (11.55,#1-0.12) rectangle ++(2,1.14);
    \node[font=\tiny\bfseries, cBad, below] at (12.55,#1-0.12) {seed};
  }
  \node[anchor=west, font=\scriptsize\bfseries] at (-0.2,3.75) {(a) Full database: entries stored back to back};
  \strip{2.0}{\cellA}{end of entry A, B's neighbor in the file}
  \draw[decorate, decoration={brace, amplitude=3pt, mirror}, thick, cBad] (5.55,1.75) -- (11.45,1.75) node[midway, below=2pt, font=\tiny, cBad] {16 bytes read before the seed};
  \node[draw=cTare, very thick, rounded corners=1.5pt, fill=cTare!8, align=left, anchor=west] at (27,2.45)
    {filter \textbf{passes} $\rightarrow$ B reported};
  \node[anchor=west, font=\scriptsize\bfseries] at (-0.2,-0.1) {(b) Replay subset: A was not a candidate, so B's neighbor is now C};
  \strip{-2.0}{\cellC}{end of entry C, different bytes}
  \draw[decorate, decoration={brace, amplitude=3pt, mirror}, thick, cBad] (5.55,-2.25) -- (11.45,-2.25) node[midway, below=2pt, font=\tiny, cBad] {same 16-byte window, different bytes};
  \node[draw=cBad, very thick, rounded corners=1.5pt, fill=cBad!6, align=left, anchor=west] at (27,-1.55)
    {filter \textbf{rejects} $\rightarrow$ B \textbf{lost}};
\end{tikzpicture}
\caption{\textbf{Why DIAMOND cannot be replayed exactly.} DIAMOND stores database entries back to back, separated by a
single byte. Its first filter examines a window of 16 bytes before and 32 after each short matching ``seed'' without
stopping at the entry boundary (DIAMOND 2.2.5, \texttt{src/search/hamming/finger\_print.h}). A seed near the start of entry B therefore
reads the end of whichever entry precedes B. Removing that neighbor from the database changes the window and can drop
B's hit. The neighbor pad adds A back, the file\nobreakdash-order neighbor of every candidate; B's line then reappears, the check
sees a line change, and \sys refuses (R5). Each of the six lines DIAMOND lost was restored by exactly one such neighbor.}
\label{fig:diamond}
\end{figure*}

\begin{table}[t]
\caption{Safety controls under the shipped check. Every case known to be wrong was refused, the correct case passed,
and no run produced wrong output.}
\label{tab:safety}
\centering\small
\setlength{\tabcolsep}{3pt}
\begin{tabular}{@{}lll@{}}
\toprule
Control & Expected & Result \\
\midrule
DIAMOND, 6 lines lost & refuse & \dec{refuse} \\
DIAMOND, 22 lines differ & refuse & \dec{refuse} \\
5\% candidates dropped & refuse & \dec{refuse} \\
HMMER genome (correct) & pass   & pass, exact \\
DIAMOND, fresh sample & no wrong output & \dec{safe-refuse} \\
BLAST+ boundary & refuse & \dec{refuse} (R1) \\
\bottomrule
\end{tabular}
\end{table}

\textbf{A failure, and what it revealed.} The check's design came from a failure. When we declared DIAMOND's database\nobreakdash-size
option (\texttt{-{}-dbsize}) as a pin, the replay on 300 proteins lost 6 of 31{,}377 stock lines while every other line
was byte\nobreakdash-identical, and our first check, which used only random pads, passed it. Under our pre\nobreakdash-registered rule, that was
an \dec{unsafe} result and invalidated that check. To find the cause, we added left\nobreakdash-out entries back in halves until each
missing line returned. Every one of the six lines was restored by exactly one entry: the candidate's neighbor in the
database file. \Cref{fig:diamond} shows the mechanism in DIAMOND's source: its first filter reads bytes across the
boundary into the preceding entry. DIAMOND's results therefore depend on the order of the database. A random pad contains one particular neighbor with probability 0.049, so three random pads miss it
86\% of the time, which is why the first check passed.

Five other Swiss\nobreakdash-Prot file orders on the same 2{,}000 queries changed 322 of 210{,}460 stock lines (173 missing, 148 extra, 1 value); 321 of those alignments sat inside the filter window of an entry boundary.

\textbf{The shipped check.} Adding the neighbor pad closed this gap (\cref{tab:safety}). On a fresh sample of 2{,}000
proteins, the replay differed from stock in 22 of 227{,}223 lines (20 missing, 2 extra), (0.0097\% of lines), and the check failed on all five pads: a true \dec{safe-refuse}. A second fresh DIAMOND sample was also refused. BLAST+ was refused earlier, at contract
inference: one of 15{,}125 aligned lines has a percent identity (column 2) that matches no normalizer, so R1 fired. That line is a different alignment, measured below. In \cref{tab:safety} the correct HMMER case passed; across all
HMMER updates, 6 of 63 were refused needlessly (\cref{sec:eval-exact}).

BLAST+ is still refused at contract inference (R1): unpinned probes print a different alignment on some rows, so percent
identity fits no column rule. A pinned pilot was byte\nobreakdash-identical to stock; we have not moved the probe under the pin, so
BLAST+ does not ship.

\begin{figure}[t]
\centering
\includegraphics[width=\linewidth]{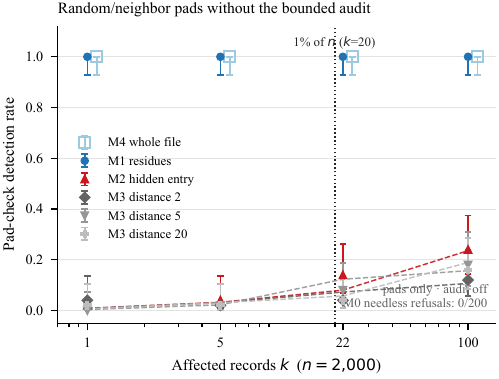}
\caption{\textbf{Pads alone are not the audit.} Detection rate of the shipped pad check on a synthetic scorer
($n{=}2{,}000$), Wilson 95\% intervals. Dashed curves: random\nobreakdash-pad formula (\cref{eq:miss}).
Vertical dotted line: $k{=}20$ (1\% of $n$), the prevalence the $s{=}299$ audit is sized for---\textbf{that audit is
off in this figure}. Labels are synthetic pad mechanisms (M-*), not the coupling taxonomy C0--C4 of \cref{sec:model}.
Residue/whole\nobreakdash-file faults are caught; hidden\nobreakdash-entry and distance faults mostly are not.
Clean controls: 0/200 needless refusals. Whole\nobreakdash-file markers offset slightly right so both series at 1.0 stay visible.
\cref{eq:miss} assumes the sampled records and pad\nobreakdash-inclusion mechanism described by $h$ and $q$; one shared
responsible entry is not the general case for arbitrary record\nobreakdash-specific dependencies.}
\label{fig:adversarial}
\end{figure}

\textbf{Limits and the bounded audit.} Random pads are weak against non\nobreakdash-neighbor coupling (\cref{eq:miss}, \cref{fig:adversarial}):
on a synthetic scorer (1{,}400 trials), file\nobreakdash-wide coupling was always refused, but hidden or distant coupling missed on 741 of
800 trials. With the earlier audit, 489 of those 800 still shipped; with the bounded audit ($s{=}299$), where at least 1\% of
records were affected only 3 of 400 shipped (vs.\ 127 of 400 earlier), within the 5\% bound, and all 200 clean trials
shipped. The historical bounded audit costs $\approx$6.4\% of a stock run (vs.\ 2.1\% earlier).
\textbf{Final account of the guarantee.}
Under C0/C1 with (D)--(O), (C), and (I) in \cref{sec:model}, \cref{prop:exact} gives an exact merge under the contract
(conditional exactness).
Observed C2--C4 trigger refuse.
Separately, for hidden coupling inside a stated audit population $A$, a uniform sample of $s{=}299$ with a valid oracle
misses faults that affect at least 1\% of $A$ with probability $\le 5\%$ (\cref{eq:audit-s}).
That is the runtime assurance; it is not a bound on shipping any incorrect output.
The historical candidate\nobreakdash-bearing adversarial numbers above use $A{=}\{$records with $\ge$1 candidate$\}$; the
headline timing attaches the same $s{=}299$ rule to $A{=}R$ (all records) and is reported in \cref{tab:config}.

\begin{table}[t]
\caption{Audit configurations and their sampling populations. Each historical policy retains its own freeze and timing evidence. The all-record row is the locked MEASURED\_PROTOCOL\_v3 campaign (nine genomes, three repeats; one wall-quality FLAGGED genome excluded from the median).}
\label{tab:config}
\centering\footnotesize
\setlength{\tabcolsep}{3pt}
\begin{tabular}{@{}p{1.55cm}p{2.05cm}p{3.75cm}@{}}
\toprule
Policy & Population & Median [95\% CI] \\
\midrule
Bernoulli & $\ge$1 candidate & $4.24$ [$4.17$, $4.68$] \\
Bounded & $\ge$1 candidate & $\HeadlineBoundedMedian$ [\HeadlineBoundedCIlow, \HeadlineBoundedCIhigh] \\
All-record & all records & $\HeadlineMedian$ [\HeadlineCIlow, \HeadlineCIhigh] \\
\bottomrule
\end{tabular}
\par\smallskip\raggedright
Bernoulli: $p{=}0.02$, floor 20; jobs 13208152/13227819.
Bounded: $s{=}299$, W13 freeze, job 13260162.
All-record: $s{=}299$; $n{=}\HeadlineN$ unflagged (\FlaggedCount\ flagged); sensitivity $\SensitivityAllGenomes\times$; C/B $\CostOfCheckingMedian\times$ (\cref{tab:allrecord-genomes}).
With a valid oracle, each bounded policy misses $\ge 1\%$ affected in its population with probability $\le 5\%$.
\end{table}

\begin{table}[t]
\caption{Per\nobreakdash-genome all\nobreakdash-record $S_g$ and $C/B$. A\nobreakdash-spread $=(\max/\min{-}1)$ on stock repeats.
$^\dagger$FLAGGED (excluded from headline).}
\label{tab:allrecord-genomes}
\centering\scriptsize
\setlength{\tabcolsep}{2.5pt}
\begin{tabular}{@{}lccc@{}}
\toprule
Genome & $S_g$ & $C/B$ & A\% \\
\midrule
\texttt{GCF\_002302335.1} & \SgA & \CBA & \SpreadA \\
\texttt{GCF\_009433135.1} & \SgB & \CBB & \SpreadB \\
\texttt{GCF\_009948655.1} & \SgC & \CBC & \SpreadC \\
\texttt{GCF\_022559285.1} & \SgD & \CBD & \SpreadD \\
\texttt{GCF\_030347155.1} & \SgE & \CBE & \SpreadE \\
\texttt{GCF\_049949335.1}$^\dagger$ & \SgF & \CBF & \SpreadF \\
\texttt{GCF\_054388855.1} & \SgG & \CBG & \SpreadG \\
\texttt{GCF\_055556535.1} & \SgH & \CBH & \SpreadH \\
\texttt{GCF\_061234675.1} & \SgI & \CBI & \SpreadI \\
\bottomrule
\end{tabular}
\end{table}

A disclosed mid\nobreakdash-run rule change (candidate\nobreakdash-vs\nobreakdash-pad for dual\nobreakdash-role entries) made the check more permissive; we reran every control and every known\nobreakdash-bad case still refused. Option search finds size pins (e.g.\ HMMER \texttt{-Z}) but not reporting thresholds; a missing pin is R2, never a guess.

\subsection{Q4: Comparison with Existing Tools}
\label{sec:eval-baselines}

\begin{table}[t]
\caption{Matched \textbf{first run $\neq$ update} cost on one exclusive EPYC 7542 node (500 proteins, Pfam\nobreakdash-A 30{,}134 models,
$n{=}4$ paired): $\rho = T/T_{\mathrm{stock}}$ (mean, 95\% $t$-interval). With $n{=}4$, INCR's interval is wide
([0.87, 2.13]); do not remember only ProcessCache's $14.49\times$. ``New release'' is what happens when the
reference database changes. Riker was blocked on this cluster (\texttt{PTRACE\_GET\_SYSCALL\_INFO}, errno 5); a laptop
build measured $1.62\times$ [1.53, 1.71] on the same protein count.}
\label{tab:baselines}
\centering\small
\setlength{\tabcolsep}{4.5pt}
\begin{tabular}{@{}lcc@{}}
\toprule
System & New release & First\nobreakdash-run $\rho$ \\
\midrule
INCR~\cite{xie2026incr}                        & reruns everything & $1.50\times$ [0.87, 2.13] \\
ProcessCache~\cite{shiptoski2023processcache}  & reruns everything & $14.49\times$ [14.06, 14.92] \\
Riker~\cite{curtsinger2022riker}               & reruns everything & blocked (ptrace) \\
\sys                                           & replays changed part & $4.20\times$ [4.17, 4.24] \\
\bottomrule
\end{tabular}
\end{table}

\textbf{General\nobreakdash-purpose caches.} On one exclusive EPYC 7542 node and one workload (\cref{tab:baselines}),
INCR's first run was $1.50\times$ stock (95\% CI 0.87--2.13), ProcessCache $14.49\times$ (14.06--14.92), and \sys
$4.20\times$ (4.17--4.24); Riker was blocked (ptrace). Command\nobreakdash-keyed caches re\nobreakdash-execute on a new genome and
on a new release, so a release is outside their design.
On whole genomes under the historical Bernoulli audit, first\nobreakdash-run cost is a median $1.32\times$ stock (break\nobreakdash-even
0.26 updates); the headline update still replays only the changed part (\cref{sec:eval-time}).
Sharing one fill's contract across cold genomes stays above $1\times$ stock because each genome still pays its own full
search (separate from the release\nobreakdash-update headline; all shared\nobreakdash-contract outputs were exact under the contract).

\begin{figure}[t]
\centering
\includegraphics[width=\linewidth]{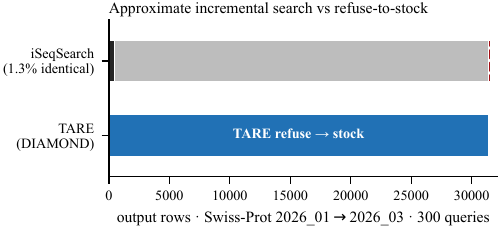}
\caption{\textbf{Approximate incremental search vs refuse\nobreakdash-to\nobreakdash-stock.} iSeqSearch's incremental DIAMOND after a
Swiss\nobreakdash-Prot update vs a full run. Of shared rows: 407 byte\nobreakdash-identical (1.3\%), 30{,}968 same hit with a different
value, 2 different alignments; plus 133 stale hits to deleted sequences. \sys refuses DIAMOND and dumps stock
(byte\nobreakdash-identical, no speedup)---a boundary case, not a HMMER result.}
\label{fig:idiamond}
\end{figure}

\textbf{Approximate incremental search.} iSeqSearch's incremental DIAMOND~\cite{yoo2025iseqsearch} on 300 proteins (Swiss\nobreakdash-Prot 2026\_01$\rightarrow$2026\_03) matched only 407 of 31{,}377 shared lines byte\nobreakdash-for\nobreakdash-byte (1.30\%) and retained 133 hits to deleted sequences (\cref{fig:idiamond}); \sys refuses and returns stock. A second locked attempt to time that refuse path failed its pre\nobreakdash-registered control (scorer \texttt{tare\_refused} false) while every timed \sys arm recorded \dec{refuse} with stock\nobreakdash-identical output; we state that as a limitation and do not claim an iSeqSearch speedup.

\begin{table*}[t]
\caption{Prior work on the problem of updating annotations after a database release. ``Unmodified'': runs the program as
is. ``Release\nobreakdash-aware'': reuses work when the database changes. ``Identical'': output equals a full run byte for byte.
``Per\nobreakdash-tool code'': requires code or formulas written for each program. $^\dagger$\sys's pins, side files, groups, and
reporting thresholds are declared as data on the command line; its core has no program\nobreakdash-specific code.}
\label{tab:prior}
\centering\footnotesize
\setlength{\tabcolsep}{3.1pt}
\begin{tabular}{@{}llcccc>{\raggedright\arraybackslash}p{3.6cm}@{}}
\toprule
System & Grain & Unmodified & Release\nobreakdash-aware & Identical & Per\nobreakdash-tool code & Limitation for this problem \\
\midrule
Riker, ProcessCache, INCR~\cite{curtsinger2022riker,shiptoski2023processcache,xie2026incr} & command / chunk & yes & no & yes & no & a release reruns everything \\
Nextflow, Snakemake~\cite{ditommaso2017nextflow,molder2021snakemake} & task & yes & no & yes & no & task inputs change \\
iBLAST~\cite{dash2021iblast} & appended sequences & yes & growth only & rescaled & yes & BLAST\nobreakdash-specific E-value formula \\
iSeqSearch~\cite{yoo2025iseqsearch} & appended sequences & yes & growth only & rescaled & yes & 1.3\% of lines identical; stale hits \\
GeStore~\cite{pedersen2014gestore} & plugin & yes & yes & plugin\nobreakdash-defined & yes & admin plugins; HMMER input\nobreakdash-only; Pfam marks all updates \\
Precomputed lookups~\cite{jones2014interproscan,cantalapiedra2021eggnog} & query sequence & yes & no & yes & no & keyed by query, fixed release \\
KumQuat~\cite{shen2022kumquat} & input stream & yes & no & yes & no & no database argument \\
IVM, self\nobreakdash-adjusting~\cite{blakeley1986views,acar2003selective} & operator & no & yes & yes & n/a & computation must be rewritten \\
mpiBLAST~\cite{darling2003mpiblast} & database segment & yes & no & yes & yes & parallel search of one release \\
\textbf{\sys} & \textbf{record $\times$ entry} & \textbf{yes} & \textbf{yes} & \textbf{under contract} & \textbf{no}$^\dagger$ & \textbf{refuses observed unsupported coupling; audit is probabilistic} \\
\bottomrule
\end{tabular}
\end{table*}

\textbf{GeStore, by axis.}
GeStore already increments unmodified genomics tools, including HMMER, via administrator plugins (parser, input
generator, optional merger)~\cite{pedersen2014gestore}.
Both leave the binary alone; the axes differ.
\textit{Knowledge:} GeStore's METApipe plugins were hundreds of Java lines; \sys declares pins and side files as data
(\cref{sec:impl}).
\textit{Correctness:} GeStore's BLAST plugin rescales E-values at merge; its HMMER plugin is input\nobreakdash-only, with no
refuse gate---correctness is the plugin author's.
\sys reprints under a pin and refuses on pad/audit disagreement; identity is conditional on \cref{prop:exact}.
\textit{Performance:} GeStore's HMMER path marked all Pfam updates significant; \sys selects candidates and times the
all\nobreakdash-record audit (\cref{sec:eval-time}).
We did not re\nobreakdash-implement GeStore, so this is mechanistic, not a matched bake\nobreakdash-off.

\section{Discussion}
\label{sec:discussion}

\textbf{Semgrep boundary.} On the measured run (\texttt{-j 16}, $r{=}2$), updates shipped stock\nobreakdash-identical except volatile
JSON \texttt{time}, but $S{\approx}0.50$ because \texttt{-{}-verify full} is itself a stock scan ($f_\Delta{=}0.0049$).
With HMMER (pays) and DIAMOND/BLAST+ (refused), Semgrep is exact under the contract, not profitable---a boundary, not a
third successful acceleration.

\textbf{Threats to validity.}
\textit{Breadth:} HMMER is primary; Infernal shares lineage with a reduced long\nobreakdash-DNA check; DIAMOND/BLAST+ refuse;
Semgrep is exact but slower---not a third acceleration (\cref{sec:eval-safety}).
We claim exposed\nobreakdash-setting coupling, not most search programs.
\textit{Release/timing:} headline is 38.1$\rightarrow$38.2 (13.4\%); \cref{fig:pairs} spans 2.0\%--93.1\%; laptop numbers are indicative.
\textit{Check:} below 1\% prevalence, hidden coupling can ship (\cref{tab:config}).

\textbf{Closed as inconclusive: isolation from a competent expert.}
Command caches predictably rerun when the database file changes; the decisive baseline is a release\nobreakdash-aware expert
with the same declared pin and sound candidate selection.
A shell script that only reimplements the batch grid is \sys without gate/check/audit, not that expert.
The locked five\nobreakdash-genome draw (exclusive EPYC 7542; seeds 20261242/20261249) is complete as evidence:
three genomes scored, with median $S_g^{\mathrm{tare}}{=}5.32$ and $S_g^{\mathrm{expert}}{=}5.96$ ($R_g{=}0.866$);
two genomes (\texttt{GCF\_019969525.1}, \texttt{GCF\_040271735.2}) refused at fill on duplicate domain\nobreakdash-table row keys,
so they contribute no paired $S_g$. Status \texttt{inconclusive}; \texttt{locked\_claim=false}. We do not claim a
speedup over expert composition, and we do not treat a redraw as unfinished work on this endpoint.
Campaign C/B $\CostOfCheckingMedian\times$ is the cost of the certificate on the unflagged all\nobreakdash-record set, not
evidence that packaging known techniques beats a careful human script.

\section{Related Work}
\label{sec:related}

\textbf{What is new.} Command caches and workflow managers~\cite{curtsinger2022riker,shiptoski2023processcache,xie2026incr,ditommaso2017nextflow,molder2021snakemake}
rerun when the reference file changes.
Caruca mines specifications of opaque components rather than caching command results~\cite{lamprou2025caruca}.
iBLAST/iSeqSearch rescale append\nobreakdash-growth~\cite{dash2021iblast,yoo2025iseqsearch}; GeStore~\cite{pedersen2014gestore}
is compared in \cref{sec:eval-baselines}.
mpiBLAST~\cite{darling2003mpiblast} contributes the pin; \sys adds candidate selection, batched replay, and
check\nobreakdash-or\nobreakdash-refuse with a prevalence audit.
Black\nobreakdash-box splitters and IVM~\cite{shen2022kumquat,vasilakis2021pash,blakeley1986views,acar2003selective} do not
split a size\nobreakdash-coupled database argument of an unmodified binary.
Probes borrow metamorphic testing~\cite{chen1998metamorphic,segura2016survey}.
HMMER flag semantics (\texttt{-Z}, \texttt{-{}-domZ}) follow the HMMER~3.4 User's Guide~\cite{eddy2023hmmer34guide};
the 2011 research article~\cite{eddy2011hmmer} is the algorithmic reference.

\section{Conclusion}
\label{sec:conclusion}

For programs whose coupling lies in exposed settings, pinning lets an unmodified binary reprint under a stated contract
while searching only what changed; pads and a prevalence audit refuse when that contract is falsified (as for DIAMOND).
Conditional exactness (\cref{prop:exact}) and the $s{=}299$ audit are separate: the latter does not cover arbitrary rare
faults.
On Pfam 38.1$\rightarrow$38.2 under the locked all\nobreakdash-record audit, median $S_g{=}\HeadlineMedian\times$
(\cref{tab:config}).
Open: expert isolation, and a third independent application under the full check.

\FloatBarrier
\paragraph{Availability.}
Versioned artifact (code, protocols, committed results, figure scripts, campaign\_v3 analysis and
\texttt{verify\_campaign\_v3\_headline.py}):
\url{https://github.com/thatrandomasiandev/tare/releases/tag/artifact-20261101}.
DIAMOND boundary analysis cites revision 2.2.5, \texttt{src/search/hamming/finger\_print.h}.\label{paper:body-end}

\clearpage
\bibliographystyle{plain}
\bibliography{refs}

\end{document}